\documentclass[a4paper, amsfonts, amssymb, amsmath, reprint, showkeys, nofootinbib, superscriptaddress, twoside, longbibliography]{revtex4-1}
\usepackage[english]{babel}
\usepackage[utf8]{inputenc}
\usepackage[colorinlistoftodos, color=green!40, prependcaption,textsize=small]{todonotes}
\usepackage{amsthm}
\usepackage{mathtools}
\usepackage{physics}
\usepackage{xcolor}
\usepackage{graphicx}
\usepackage[left=23mm,right=13mm,top=35mm,columnsep=15pt]{geometry} 
\usepackage{adjustbox}
\usepackage{placeins}
\usepackage[T1]{fontenc}
\usepackage{lipsum}
\usepackage{csquotes}
\usepackage{xcolor}
\usepackage{hyperref}
\hypersetup{
    colorlinks,
    linkcolor={red!50!black},
    citecolor={green!50!black},
    urlcolor={blue!80!black}
}\usepackage{soul}
\usepackage{changes}
\usepackage{xargs}                      
\usepackage{xcolor}
\definecolor{alicecolor}{RGB}{245, 111, 111}
\definecolor{bobcolor}{RGB}{160, 180, 241}
\definecolor{sourcecolor}{RGB}{200, 200, 200}
\usepackage{tikz}
\usetikzlibrary{shapes.misc,arrows}

\begin{document}

    \title{The Sound of Entanglement}
    
\author{Enar de Dios Rodríguez}
    \affiliation{Kunstuniversität Linz, Hauptplatz 8, 4010 Linz, Austria}
    \affiliation{IFK, Internationales Forschungszentrum Kulturwissenschaften Kunstuniversität Linz in Wien, Reichsratsstraße 17, 1010 Wien, Austria}
\author{Philipp Haslinger}
    \affiliation{Vienna Center for Quantum Science and Technology, Atominstitut, Technische Universität Wien, Vienna, Austria}
 \affiliation{University Service Centre for Transmission Electron Microscopy, Technische Universität Wien,
Vienna 1040, Austria}
\author{Johannes Kofler}
    \affiliation{Department of Quantum Information and Computation at Kepler (QUICK), Johannes Kepler University Linz, Austria}
\author{Richard Kueng}
    \affiliation{Department of Quantum Information and Computation at Kepler (QUICK), Johannes Kepler University Linz, Austria}
\author{Benjamin Orthner}
    \thanks{corresponding authors:
        \href{mailto:alexander.ploier@jku.at}{alexander.ploier@jku.at}\\
        \phantom{corresponding authors: }\href{mailto:benjamin.orthner@tuwien.ac.at}{benjamin.orthner@tuwien.ac.at}}
    \affiliation{Vienna Center for Quantum Science and Technology, Atominstitut, Technische Universität Wien, Vienna, Austria}
\author{Alexander Ploier}
    \thanks{corresponding authors:
        \href{mailto:alexander.ploier@jku.at}{alexander.ploier@jku.at}\\
        \phantom{corresponding authors: }\href{mailto:benjamin.orthner@tuwien.ac.at}{benjamin.orthner@tuwien.ac.at}}
    \affiliation{Department of Quantum Information and Computation at Kepler (QUICK), Johannes Kepler University Linz, Austria}
\author{Martin Ringbauer}
    \affiliation{Institute for Experimental Physics, University of Innsbruck, Austria}
\author{Clemens Wenger}
    \affiliation{Institute for Composition, Conducting and Computer Music (IKD), Bruckner University Linz, Austria}

\date{\today} 

\begin{abstract}
The advent of quantum physics has revolutionized our understanding of the universe, replacing the deterministic framework of classical physics with a paradigm dominated by intrinsic randomness and quantum correlations. This shift has not only enabled groundbreaking technologies, such as quantum sensors, networks and computers, but has also unlocked entirely new possibilities for artistic expressions. In this paper, we explore the intersection of quantum mechanics and art, focusing on the use of quantum entanglement and inherent randomness as creative tools. Specifically, we present \textit{The Sound of Entanglement}, a live musical performance driven by real-time measurements of entangled photons in a Bell test. By integrating the measured quantum correlations as a central compositional element and synchronizing live visuals with experimental data, the performance offers a unique and unrepeatable audiovisual experience that relies on quantum correlations which cannot be produced by any classical device. Through this fusion of science and art, we aim to provide a deeper appreciation of quantum phenomena while expanding the boundaries of creative expression.

\end{abstract}
\keywords{quantum physics, entanglement, randomness, aleatoric music, quantum art}

\maketitle

\newpage
\section{Introduction}\label{ch:Intro}



The advent of quantum physics fundamentally changed the way we understand the world around us. While the universe once was thought of as a deterministic clockwork, it is now becoming clear that intrinsic randomness and quantum correlations reign supreme in the microscopic world. This paradigm shift allowed us to break out of the limits of classical physics and develop new technologies from quantum sensors over unconditionally secure quantum communication protocols to quantum computers that can outperform even the best supercomputers at certain, albeit specific, tasks. These new technological developments, in turn, open new opportunities for arts and science. Quantum concepts and quantum computers are already being used for the composition of paintings~\cite{crippa2024a} and music through the sonification of quantum computations or subroutines~\cite{clemente2022newdirectionsquantummusic, itaborai_2023_10206731, Weaver2022, Hamido2023, yamada_2023_10206508,QuantumMusic}. Sonification of scientific data has a long history in different fields~\cite{zanella2022,candey2006sonification, diazmerced2013sonification, quinn2001sonification} and can be a powerful way to understand and experience physics.

\begin{figure}[t]
    \centering
    \includegraphics[width=\linewidth]{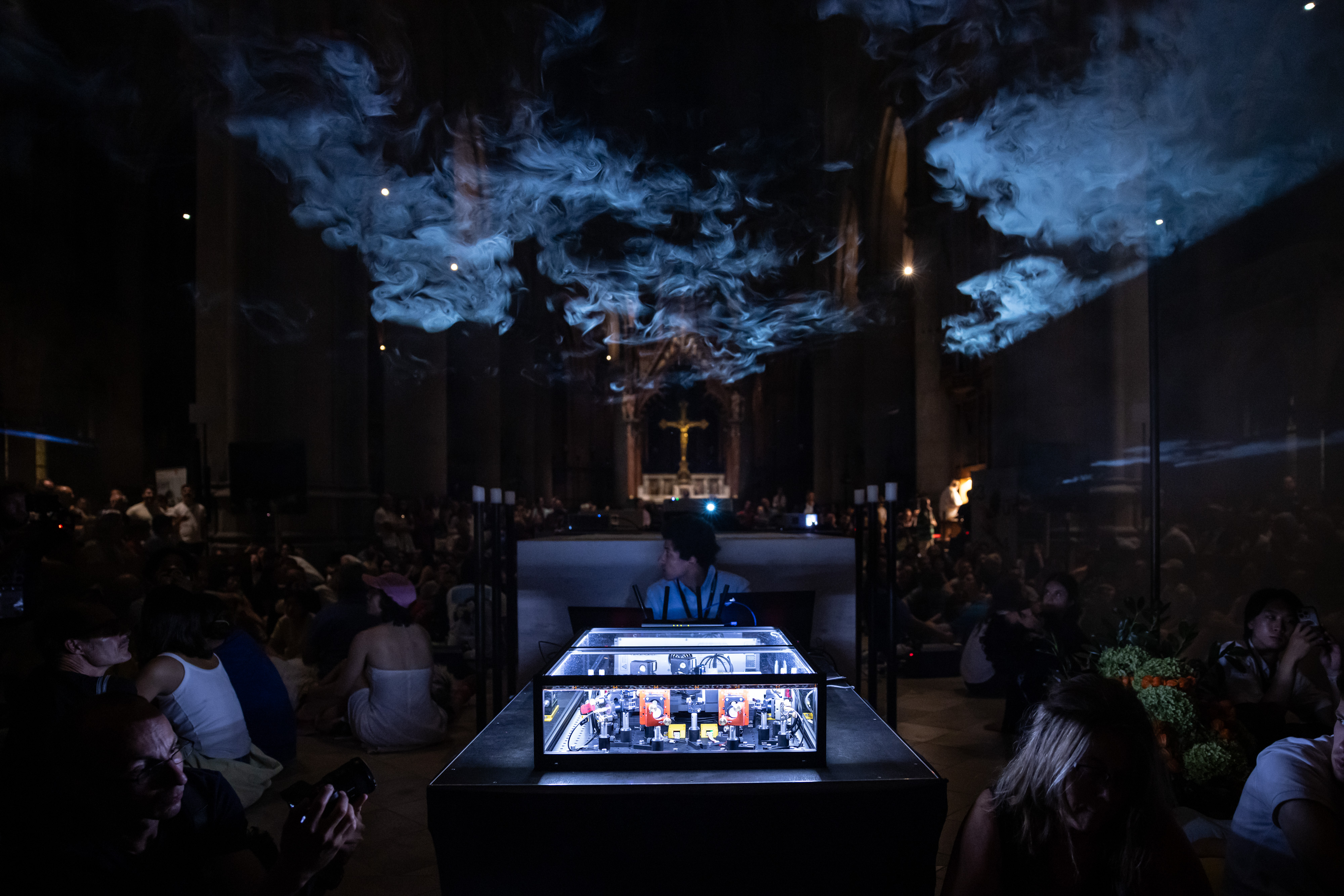}
    \caption{\emph{Sound of Entanglement world premiere}. A picture taken during the world premiere of \textit{The Sound of Entanglement} in the New Cathedral in Linz, Austria, as a part of the opening act for the Ars Electronica Festival on September 4th, 2024, which coincided with Anton Bruckner's 200th birthday. The Bell setup, as an integral part to our performance, sits in the middle of the church and is highlighted by some additional light elements. The projectors are positioned such that the audience is sitting in the visual projection. \textcopyright Ars Electronica}
    \label{fig:worldpremiere}
\end{figure}

Complementary to traditional sonification of experimental data, the underlying fundamental physical effects can create new creative expressions. Here, we focus particularly on quantum entanglement, which, as Erwin Schrödinger put it, is ``not [...]~\emph{one} but rather \emph{the} characteristic trait of quantum mechanics, the one that enforces its entire departure from classical lines of thought.''~\cite{Schrodinger1935Steering} In our project \emph{The Sound of Entanglement}, the fundamental randomness and stronger-than-classical correlations which arise from entangled quantum systems are created live on-stage and used as a central creative element. The project was first realized as a special musical and visual composition titled \textit{BruQner} for the opening of the Ars Electronica Festival 2024 and the Anton Bruckner year 2024 (see Fig.~\ref{fig:worldpremiere}).

\subsection{Aleatoric Music}
Randomness and chance have been explored by composers as creative tools for centuries, with improvisation being a fundamental aspect of musical practice long before the 18th century. In fact, improvisation has been a cornerstone of many musical traditions, from the extemporaneous embellishments of medieval and Renaissance performers to the spontaneous creation of melodies in oral traditions around the world. By the 18th century, randomness began to take on more formalized roles in Western music. For example, Mozart is said to have devised a dice game for his pupils, allowing them to create waltzes ``without understanding anything about music or composition''~\cite{mozart_wuerfelspiel}.
From the 1950s onwards, the term ``aleatoric music'' gained popularity as a way to describe the diverse musical approaches that embrace chance as a fundamental element in the creative process. Among these works, John Cage's \textit{4'33''}~\cite{cage_433} stands out as one of the most enigmatic pieces in Western cultural memory. In this piece, the composer only specifies the length of time during which the performer should perform silence. This formal requirement thus makes random ambient sounds, rather than intentionally played notes, the protagonists of the composition, making the acoustic experience entirely unpredictable and unique with each rendition~\cite{schulze2000aleatorische}.

Aleatoric music also incorporates elements of combinatorics and improvisation, which are not only central to this genre but are also characteristic of numerous popular musical styles, including jazz and contemporary electronic music. These connections highlight the enduring role of chance and spontaneity in music across both historical and modern contexts. In particular, the subjective randomness introduced by human performers—through their individual decisions, interpretations, and the inherent variability of playing an instrument—adds a unique layer of unpredictability that intertwines with the structured randomness of aleatoric techniques.

\subsection{Quantum Aleatoric Music}

\begin{figure*}
    \centering
    \includegraphics[width=0.9\linewidth]{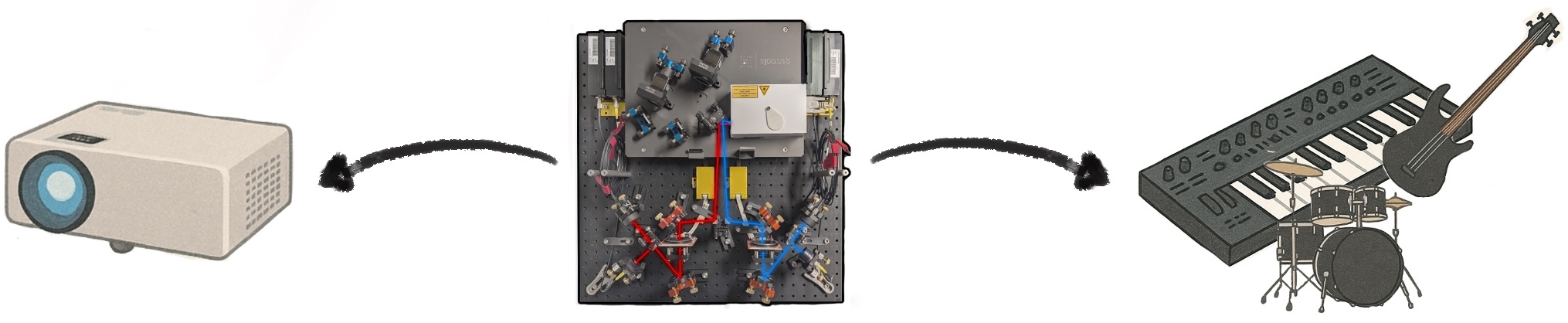}
    \caption{\emph{Sketch layout of the project components.} The experimental Bell setup in the middle, where data detection happens live on-stage. This quantum data of non-classical correlations is then transferred to the projector, here on the left, and to the musicians, symbolized through three instruments. The communication happens via laptops and LAN connections (excluded in the sketch).}
    \label{fig:performance_layout}
\end{figure*}

Contrary to the examples mentioned above, \textit{The Sound of Entanglement} introduces the notion of ``quantum aleatoric music'' which, for the first time, not only brings the objective, non-deterministic randomness of individual quantum events but also the non-classical correlations arising from entanglement into a live musical performance. In order to achieve this, we use an experimental setup which tests whether the correlations between measurements on entangled particles can be explained by local hidden variables, i.e.\ hypothetical variables that supposedly determine the outcomes of quantum measurements while obeying the rules of classical physics, or if they follow the predictions of quantum mechanics. This is also called a Bell test~\cite{bellEinsteinPodolskyRosen1964}. We use such a Bell setup and employ the entangled photons to conduct the music. To further enrich the audience's experience, the musical performance is accompanied by live visuals synchronized with the real-time measurements of entangled photons, effectively integrating the experimental data as the video jockey (VJ) of the project. Due to this experimental setup taking over the role of the conductor of the musical and visual performance, we introduce the notion of it being the \textit{quantum conductor}  (see Fig.~\ref{fig:performance_layout}).

In the spirit of the pioneers of aleatoric music, our aim is to not only develop music that allows for objective chance and real indeterminacy, but music that also confronts its interpreters with the counterintuitive and uncanny consequences of quantum physics. The aforementioned methodologies and characteristics of classical aleatoric music are expanded upon by placing them in the environment of non-classical correlations. 

\subsection{Quantum Randomness and Entanglement}

In the realm of quantum mechanics, phenomena often defy the intuitive grasp of our classical world view. Among these enigmatic features which topple the framework of our classical physics are the inherent randomness \cite{kofler2010randomness} of individual quantum events, the entanglement \cite{EinsteinPodolskyRosen1935} between quantum systems, and Bell's theorem \cite{bellEinsteinPodolskyRosen1964}.

Unlike classical physics, where systems (particles) evolve in a deterministic manner, quantum mechanics posits that some outcomes cannot be predicted with certainty. Instead, the theory provides only probabilities for different outcomes. For example, whether a photon is reflected or transmitted at a semi-transparent mirror happens without any cause, i.e.\ non-deterministically. Instead, the theory states that both possibilities exist in a superposition until the photon is detected at one or the other side of the mirror. Similarly, the polarization of a photon may be in a superposition of horizontal and vertical polarization. 
Quantum entanglement is the phenomenon where two or more quantum systems are interlinked in such a way that a full description of the joint system is not possible by fully describing the individual systems. This is a complete break with the laws of classical physics and there is no analog to it in our everyday world. This peculiar connection between entangled quantum systems was first proposed by Albert Einstein, Boris Podolsky, and Nathan Rosen in 1935 (EPR) \cite{EinsteinPodolskyRosen1935}. As it holds true for systems which are far apart from each other, Einstein called it ``spooky action at a distance''. EPR were confident at that time that quantum mechanics is incomplete and that new, hitherto undiscovered or unappreciated, ``hidden variables'' would allow a restoration of a classical deterministic worldview.

Our classical worldview is captured by the concept of \textit{local realism} (or ``local hidden variables''), which comprises (i) \textit{realism} -- i.e.\ the worldview that systems possess their properties prior to and independent of measurement via so-called hidden variables -- and (ii) \textit{locality} -- i.e.\ the principle that no influence can propagate faster than the speed of light. Quantum entanglement challenges this local realist worldview. Bell's theorem, introduced by John S.\ Bell in 1964~\cite{bellEinsteinPodolskyRosen1964}, showed that local realism is \textit{incompatible} with the predictions of quantum mechanics. While local realism has to satisfy an inequality on the correlations obtained by measurements on distant objects, quantum entangled systems allow to violate this so-called Bell inequality. Remarkably, numerous experiments have confirmed that Bell's inequalities are indeed violated \cite{hensen2015loophole-free,Giustina-Zeilinger2015,Shalm2015LoopholeFree,CHinequality,CHSH1969}, supporting the quantum mechanical view that entangled particles are correlated in a way that cannot be explained by any classical theory.
The interplay of entanglement and the inherent randomness of quantum mechanics paints a picture of a world that is profoundly different from our deterministic classical universe. These concepts not only expand our understanding of the physical universe but also inspire new ways of thinking about art and music. The non-local connections and special correlations create novel ways in which art can transcend boundaries and explore uncharted horizons. 

\section{Related Work}

As already discussed in Section~\ref{ch:Intro}, subjective randomness has been a part of musical history since well before the 18th century. Artists gained knowledge of and started experimenting with ``quantum'' as a general concept for their art around the turn of the millenium, due to the ever-growing quantum community itself. Here, quantum is a very loose terminology. It can range from art that is inspired by quantum technology to art using quantum generated data. From quantum states describing certain combinatorial art effects to quantum particles directing artists in one way or the other. An early thought experiment in the creation of quantum music involved the generation of quantum musical tones~\cite{QuantumMusic}. The authors suggest bundling one octave of a musical scale — for example, the C major scale from C6 to B6 — into a single quantum tone. In Ref.~\cite{clemente2022newdirectionsquantummusic, itaborai_2023_10206731} two other general ideas are discussed. The first one is to turn a quantum computer into an actual musical instrument, i.e. a quantum keyboard. Quantum states are initialized, transformed by operations, and then measured. The result of this measurement is mapped to some sound feature. Another possible way is to sonify the iteration steps in a variational quantum algorithm which finds the solution to a given optimization problem.

\textit{The Sound of Entanglement}, on the other hand, goes beyond earlier approaches, as we work with the observable consequences of the fundamentally \textit{non-classical correlations} between entangled quantum systems. We create pairs of entangled photons, live and on-stage, in a so-called Bell experiment. For further details regarding the setup, we refer to Section~\ref{ch:ST}. In contrast to Ref.~\cite{QuantumMusic, itaborai_2023_10206731,clemente2022newdirectionsquantummusic}, our work brings a Bell setup live on-stage, where \textit{two} musicians translate the outcomes of measurements on entangled quantum states -- which violate Bell's inequality -- into a musical and visual performance. For a more detailed explanation, we refer to Section~\ref{ch:Results}.

\section{Setup, Bell, and Mapping}\label{ch:ST}

\subsection{Setup}

\begin{figure}[t]
\begin{center}
\includegraphics[trim=0cm 0cm 0cm 1cm,clip,width=\linewidth]{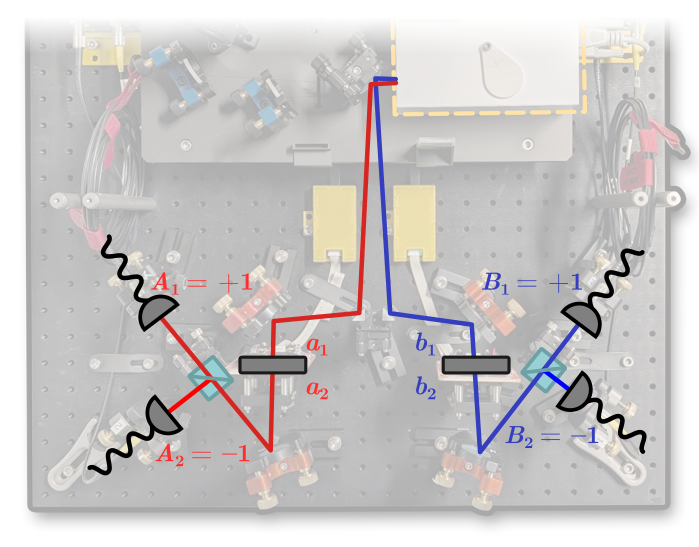}
\caption{\emph{Bell setup.} A laser source (white box) produces entangled photon pairs that are guided through mirrors and half-wave plates to two detectors: Alice (red) and Bob (blue). Both choose between two measurement settings -- $a_1,a_2$ for Alice, $b_1, b_2$ for Bob -- and upon measuring their photon get an output with the value $+1$ or $-1$.}
\label{fig:Bell_setup}
\end{center}
\end{figure}

At the heart of our setup (see Fig.~\ref{fig:Bell_setup}), which was built at the TU Wien, lies a source of polarization-entangled photon pairs. Those pairs of photons are created through a process called spontaneous parametric downconversion~\cite{SPDC}, where, in our case, a photon from a laser with a wavelength of $405$\,nm is converted, within a non-linear crystal, into a pair of photons with a wavelength of $810$\,nm each. Due to energy and momentum conservation, the two photons are strongly correlated. In addition, by choosing the right input polarization, a situation can be engineered where the photons are entangled in their polarization degree of freedom. In the process of this downconversion the following entangled Bell state is produced
\begin{eqnarray*}
    \ket{\Phi^+} = \frac{\ket{\textrm{HH}}+\ket{\textrm{VV}}}{\sqrt{2}}.
\end{eqnarray*}
Upon measurement in the horizontal (H) / vertical (V) basis, each photon will give a random result -- either it is horizontal or vertical, with 50:50 odds -- but both photons certainly show the same result. For an introduction to quantum states and entanglement, we refer to~\cite{nielsen2000quantum}.

A series of mirrors then direct these two photons to spatially separated measurement stations, traditionally referred to as \emph{Alice} and \emph{Bob}. At each measurement station, a motorized half-wave plate is used to randomly select one of two pre-programmed polarization directions to be measured with a switching time of less than $100$\,ms.

A full cycle of rotating the half-wave plates to switch to another measurement basis, measuring, and the buffer time needed for the synchronization takes around $400$\,ms between successive measurement results. After the measurement choice, the two possible polarizations are separated using a polarizing beam splitter. The respective photons are then coupled into optical fibers and sent to single-photon avalanche diodes, where the presence of a photon is converted to an electrical signal or a \emph{detector click}. From now on we refer to those electrical signals simply as \emph{click}. In order to correlate the measurement results obtained by Alice and Bob, and to make sure that they indeed originate from the same photon pair, each click is assigned a precise time stamp using a time tagging module with a resolution below $350\,$ps. Two clicks are then considered to have originated from the same photon pair, if they arrive within $1$\,ns of each other. 




\subsection{Bell's Inequality}

The source produces pairs of entangled photons. From each pair, one photon is sent to Alice, and the other photon is sent to Bob. Each party can randomly choose between two different measurements (polarization directions). Alice's options are called $a_1$ and $a_2$, while Bob's choices are $b_1$ and $b_2$. Every photon at Alice then produces a click (i.e.\ a measurement result) in one of two detectors. Alice labels her outcome $A=+1$, if her first detector clicks, and she labels her outcome $A=-1$ if her second detector fires. Similar for Bob, who calls his outcomes $B=+1$ or $B=-1$. Note that each photon in the pair is detected in just one arm after the beam splitter, so each measurement setting yields a single outcome.

Creating many photon pairs and repeating again and again the procedure of random measurement choice and outcome observation, Alice and Bob can estimate correlations between their measurement outcomes. E.g.\ how often does Alice get $A=+1$ with measurement $a_2$, when simultaneously Bob gets outcome $B=-1$ for setting $b_1$. There are $(2\times2)^2 = 16$ such combinations: $2$ outcome possibilities for each of the $2$ setting choices for each of the $2$ parties. In turn, estimating the expectation value over the outcomes produces $4$ different correlation values that can be further combined into a single number, called Bell value or $S$-value. In the Supplementary Material, we derive the CHSH inequality~\cite{CHSH1969}, a special version of Bell's inequality suitable for our setup. It states that classical systems -- i.e.\ systems that are described by local hidden variables (local realism) -- can lead to correlations between Alice's and Bob's measurements of only a certain strength, namely
\begin{align}
    S \le 2.
\end{align}
Quantum mechanically entangled systems, however, in particular our entangled photons, can violate the CHSH inequality and achieve a Bell value up to
\begin{align}
    S_{\textrm{QM}} = 2\,\sqrt{2}.
\end{align}
This means that such correlations between Alice and Bob cannot be explained by the aforementioned local realism. Any human conductor (or classical computer) can only send instructions to the two musicians Alice and Bob that follow the constraints of local realism. The instructions are the (hidden) variables which tell the musicians what to do. Our quantum mechanical setup (``quantum conductor''), however, can send ``entangled instructions'' such that the music follows strong correlations of two random outcomes that no classical computer or human could generate. This is the heart of our project, and hence the name \textit{The Sound of Entanglement}.

We note that our experiment does not close any of the loopholes which, in theory, would still allow for a local realist explanation of the observed measurement results. However, all loopholes have already been closed experimentally \cite{hensen2015loophole-free,Giustina-Zeilinger2015,Shalm2015LoopholeFree}, demonstrating that nature does not obey local realism.


\subsection{Mapping from Measurement to Art}

\begin{figure}
    \centering
    \includegraphics[width=0.95\linewidth]{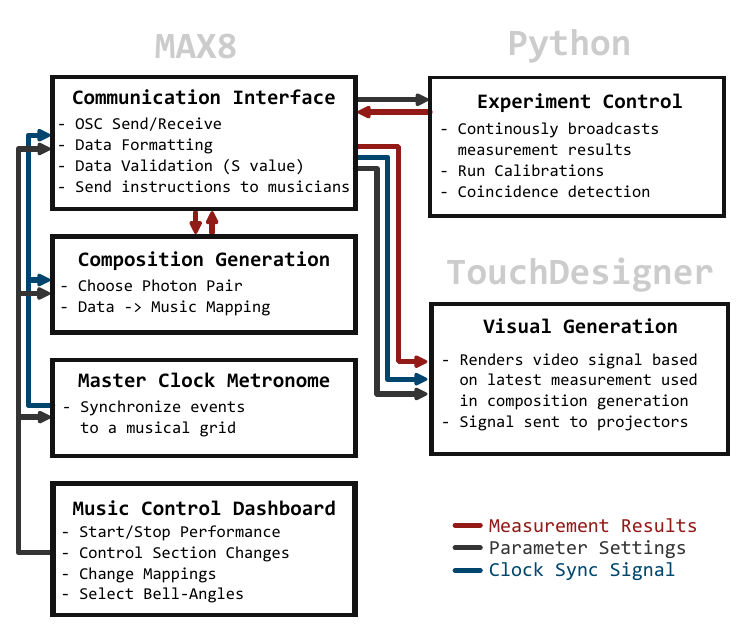}
    \caption{\emph{Software pipeline for the performance.} Detection events from the entangled-photon experiment are streamed to Experiment Control, which forwards them through a Communication Interface via OSC over Ethernet to both the Composition Generation patch and the Visual Generation engine. Arrows indicate the flow of data, control, and timing signals.}
    \label{fig:software_structure}
    \vspace{-0.5cm}
\end{figure}

Composers of quantum aleatoric music can use quantum entanglement to turn unpredictable results into correlated musical dependencies. The musical events triggered by non-classical correlations can be composed in advance. As the measurement outcomes are objectively random, the chosen musical events are in no way controllable. However, entanglement puts some control back into the hands of the composer, who can exploit the possible combinations selected by the Bell setup. Therefore, designing meaningful mappings demands a clear grasp of such an experimental setup and the theory behind it. In particular, the composer can operate the Bell setup in the regime with a Bell value $S > 2$, or deliberately tamper with the setup to shift the correlations into a regime where $S < 2$ and local realism is no longer violated. The individual parties' clicks remain (locally) random in either case, but the correlations between these outcomes change drastically. Moreover, a cleverly composed piece of music may be able to convey these differences in correlations and, in turn, make (features of) quantum entanglement audible. Achieving this has been one of the biggest challenges behind this project. More precisely, we set out to compose a mapping from Bell outcomes to musical parameters such that a shift in Bell value -- from the entangled regime to the classical limit -- is genuinely audible, turning the degree of entanglement into an intentional musical parameter. To translate the live measurements into sound, a Python daemon gathers and time-stamps the photon data, then forwards it via Open Sound Control to a Cycling ’74 Max patch. Inside Max, the mapping algorithms convert the incoming numbers into on-screen cues for the performers or control signals for virtual instruments (see Fig.~\ref{fig:software_structure}).
The same data stream drives a TouchDesigner network that generates the visuals in real time: each detected photon pair triggers animations or modulates visual parameters. Whether projected onto a screen or into stage haze, the imagery is rendered at least 30 fps and sent directly to the projector, keeping sight and sound perfectly synchronized.

\section{Results}\label{ch:Results}

As of writing this paper, we have been on tour with two very different styles of performances which we explain in the following two subsections. Both use a Bell setup as centerpiece, but differ in the way the outcomes of this experiment are mapped to musical tunes and variations, as well as the accompanying visual projections. Note that these two are merely the first two choices out of a vast ocean of mappings to explore in our quantum-to-music mapping, and we have just begun to scratch the surface. 



\subsection{BruQner}

The vision behind \textit{The Sound of Entanglement} was first realized in a special composition called \textit{BruQner} \cite{BruQnerAKM}, on September 4th and 6th, 2024, in the New Cathedral in Linz, Austria as part of the Anton Bruckner 2024 and Ars Electronica festival.
The music was played live on two organs, which we respectively call Alice and Bob. For each measurement, the quantum conductor, introduced in the beginning of this paper, chose from a set of eight precomposed musical motifs: four motifs for Alice and four for Bob. The thematic material for the sheet music is based on Anton Bruckner's organ piece \textit{Perger Präludium}\cite{BrucknerAnton1884PiCf}.\footnote{This ``open-form'' approach is characteristic of aleatoric music. E.g., in his famous piece \textit{In C} \cite{InC}, composer Terry Riley provides a set of 53 musical phrases along with rather simple playing and organization instructions for an undefined group of performers. The musical form evolves depending to the musicians decisions while playing.\cite{TerryRileyInC}}

In \textit{BruQner}, the choice of motif has been left to the quantum conductor. In addition to the software solution discussed for mapping the experiment data to music events, further challenges had to be overcome. The organists sat about 30 meters apart at the organ manuals, and they had to read the musical notes that appeared on their tablet screens only shortly before they had to be played in sync. First, we introduced a 1-bar offset to make the sheet music easier to read. In addition, WiFi transmission of graphic data and the sync signal via the OSC protocol was too inaccurate and fluctuating for live music performance (from $1\,$ms to $15\,$ms ping time). LAN cabling throughout the cathedral guaranteed a latency of about 1-2 ms, which was tolerable.

\begin{figure}
    \centering
    \includegraphics[width=0.98\linewidth]{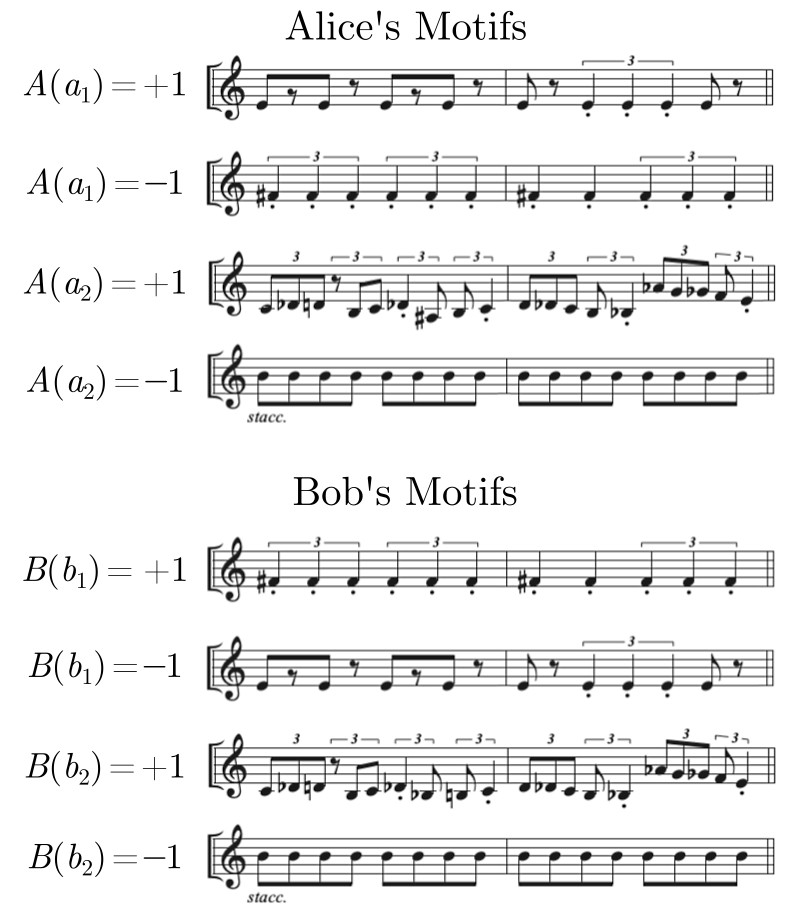}
    \caption{\emph{One possible set of 8 motifs}. In \textit{BruQner}, the quantum conductor chooses musical motifs for both musicians to play. Every time a measurement is taken, the results corresponding to one of the four musical motifs is then sent live to the musicians and queued-up for them to play. A possible set of eight motifs is shown in this figure. In this particular mapping, care was taken to ensure that non-classical correlations predominantly generate polyrhythmic motif combinations.}
    \label{fig:motifset}
\end{figure}

To vary the composition and further enlarge the space of possible music realizations, \textit{BruQner} could access $28$ different sets of motifs, each containing $4$ motifs for Alice, and $4$ for Bob, with varying bar lengths. These motifs could be recalled at different tempos synchronized to the master clock (see Fig.~\ref{fig:motifset}).

The premiere performance lasted about 18 minutes, and the measurement results showed a strong correlation between the entangled photons, with an $S$-value of approximately 2.45. In the middle of the piece, a section purely based on classically correlated data yielded an $S$-value below 2. For this part, the musical material was selected in a more arbitrary sequence to make the difference between the two approaches audible.

The visual component of the performance featured an aesthetic interplay inspired by the classical representation of quantum entanglement, symbolized by two light cones. In our interpretation, these cones evolved in real time, responding dynamically to experimental measurements and unfolding in synchrony with the musical narrative. This interplay allowed the audience to perceive the shifts between the ``quantum'' sections of the piece -- when the light cones were visible -- and the ``classical'' passages, represented by flat planes. This resulted in a unique projection each time, with no two performances ever visually identical.

\begin{figure}
    \centering
    \includegraphics[width=1\linewidth]{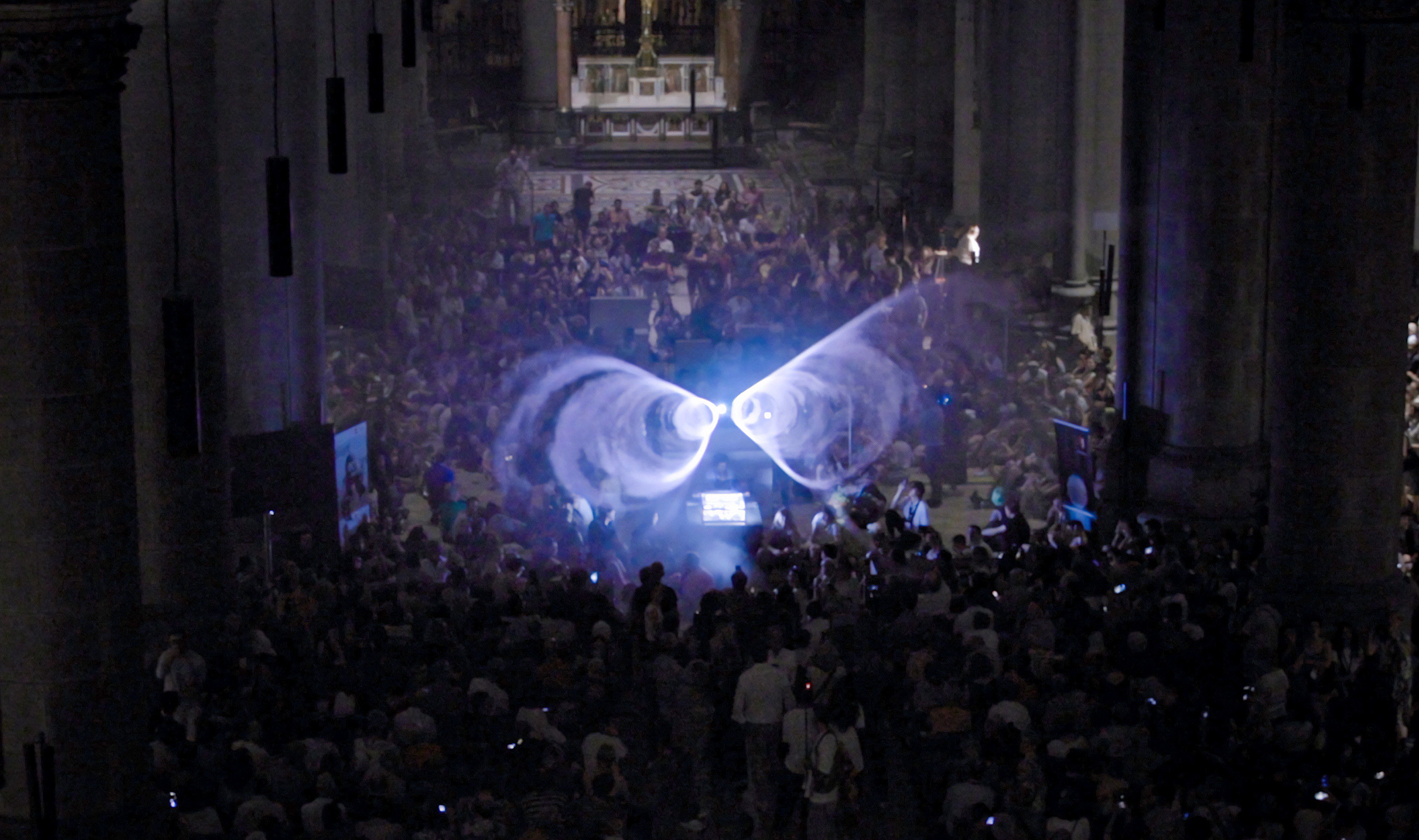}
    \caption{\emph{Still image from the documentary ``BruQner -- The Sound of Entanglement''}. The performance was filmed and a documentary was made about the days leading up to the event. Two large (and moving) light cones were projected above the audience into vapor from haze machines. These light cones were inspired by the entanglement-creation process of spontaneous parametric down-conversion, after which the two entangled photons leave the crystal along such cones.}
    \label{fig:bruQner_visuals}
\end{figure}

Departing from conventional live projection setups, where imagery is typically cast from behind the audience onto a screen in front, our design reversed this norm. Three laser projectors were positioned to face the audience, while strategically placed smoke machines close to the projectors rendered the beams of light visible in space. This configuration transformed light into a three-dimensional sculptural medium, physically inhabiting the cathedral’s vast interior. The cones of light did not simply illustrate entanglement; they embodied it -- interacting, overlapping, and surrounding the audience like living sculptures. Their shifting surfaces and intersecting edges echoed the geometric complexity of quantum phenomena, turning abstract theory into an immersive experience. We refer to Fig.~\ref{fig:bruQner_visuals} for a picture that captures these effects somewhat. A short documentary about \textit{BruQner} which follows the team of artists and scientists in the days leading up to the premiere of this live quantum music event can be watched here~\cite{BruQnerDocu}.

\subsection{8 Rooms}

The subsequent composition \textit{8 Rooms} \cite{8ROOMSAKM} also uses a Bell setup, but deviates from \textit{BruQner} as it employs different methods to map the experimental data onto musical parameters. The basis of this composition is (i) digital audio processing of live sound that is reactive to the measurements and (ii) a $2$-dimensional quantum random walk on a periodic square grid that controls the musical form. In the Supplementary Material we briefly recapitulate the mathematical model behind a $2$-dimensional random walk.

The main element of the composition is a stereo clicking sound generated by the rotation of the half-wave plates, i.e.\ coming directly from the setup. Every time the measurement basis of the Bell setup is changed and the plates move mechanically, this sound is captured: one sound for Alice and one for Bob. We timed these measurements to happen every 513 ms, enabling us to use these rotation clicks as a metronome (513 ms is 117 bpm in musical terms) and as a percussion element.

\begin{figure}
    \centering
    \includegraphics[width=0.97\linewidth]{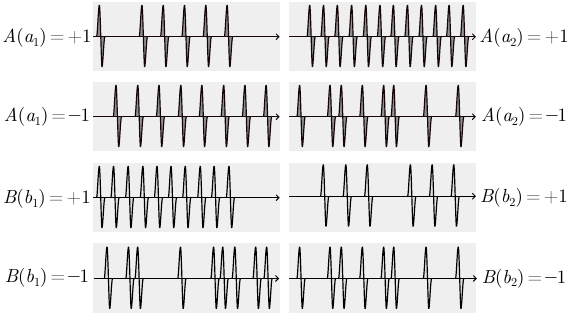}
    \caption{\emph{Types of echoes created as a basis for one rooms musical rhythm}. Alice and Bob do have four sets of click patterns each which the experiment chooses, similar to the musical motifs that have been prepared for \textit{BruQner}. They are designed in a way that every combination of echo patterns makes for a rhythmic base for the musicians to play along side.}
    \label{fig:8roomsclicks}
\end{figure}

This sound runs through the entire piece and is modified by digital audio processing. As the name of the composition, \textit{8 Rooms}, suggests, there are eight different sections, or ``rooms''. The sound aesthetics in each section are different, and each click sound picked up by the microphone is processed in real time using the experiments data. 
The measurements of Alice and Bob influence a rhythmic-pattern based on an audio delay effect, excited by the clicks of our half-wave plates. Those patterns are created akin to how \textit{BruQner} was played, i.e.\ two different delay response patterns are chosen, one for Alice and one for Bob which are then overlapped and are the baseline of the music of Room 1 (See Fig.~{\ref{fig:8roomsclicks}}).

\begin{figure}
    \centering
    \includegraphics[width=1\linewidth]{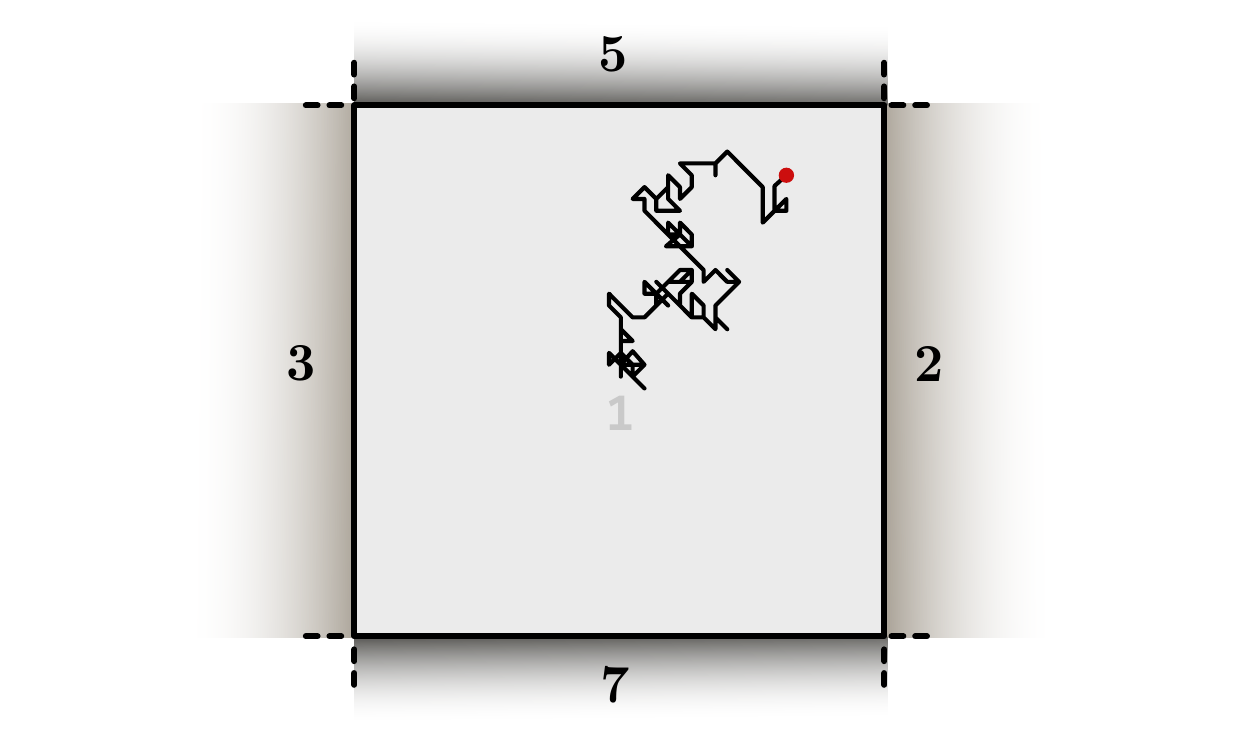}
    \caption{\emph{Example set of room compositions}. In \textit{$8$ Rooms} every room has a different musical theme. Here we show an example layout of rooms for the performance. The random walk chooses the successor to room 1 by hitting an adjacent boundary, which leads to rooms $2, 3,5,$ or $7$. When this happens the starting point is reset to the center of the new room and a new musical theme begins.}
    \label{fig:8-rooms}
\end{figure}

Switching rooms happens when the random walk, which starts at the room's center, hits one of the four boundaries of a square. Each boundary forms a door to a different room, and after switching rooms the walk is reset to the center of the new room. The composition’s path is not linear as in conventional music, where the form from start to finish is predetermined. Nor is it completely random, because Alice's and Bob's paths are strongly correlated, so the joint random walk will also tend to prefer certain exits. As an example, consider a walk that starts at the center of room 1 (see Fig.~\ref{fig:8-rooms}).
As mentioned above, the random walk starts at the center of room $1$. By hitting the boundary between room 1 and room $2$ the position is reset to the center of room $2$, and the procedure repeats. After a preset duration, all boundaries are redirected to room $9$, where the performance ends. 

The sizes of the rooms and the magnitudes of the direction vectors derived from the experimental data can be defined as compositional elements. 

\textit{8 Rooms} also brings real musicians onto the stage, who can play more freely here than the two organists in \textit{BruQner}, who strictly followed the software-generated sheet music. In \emph{8 Rooms}, the musicians follow the evolving path of the random walk and improvise over predefined, room-specific material, in a musical dialogue with the quantum-controlled clicking sounds.

Just as the digital audio signals respond to experimental measurements, the live visuals also react in real time to both the clicking sounds and data from the experiment. As explained above, the clicking sounds correspond to the rotation of the half-wave plates and thus to quantum measurement events. In the musical performance they act as the metronome, one that is continuously shaped and transformed by digital audio processing, making the beat of the piece dynamic and unpredictable. 


The visual component of \textit{8 Rooms} takes the sounds made from the half-wave plates as they change into different measurement settings as its central protagonist, rendering their histogram across a spatial visual plane. When the piece begins in silence, only two two-dimensional shapes -- constructed from fine white lines on a black background, one representing Alice and the other Bob -- are visible. As measurement events accumulate, these forms gradually transform into an undulating terrain, with each emerging peak representing a single 'mountain' on the surface, and together generating a three-dimensional landscape. In this way the histogram becomes a spatial, poetic expression of the rhythmic structure that arises from quantum uncertainty.

\begin{figure}
    \centering
    \includegraphics[width=1\linewidth]{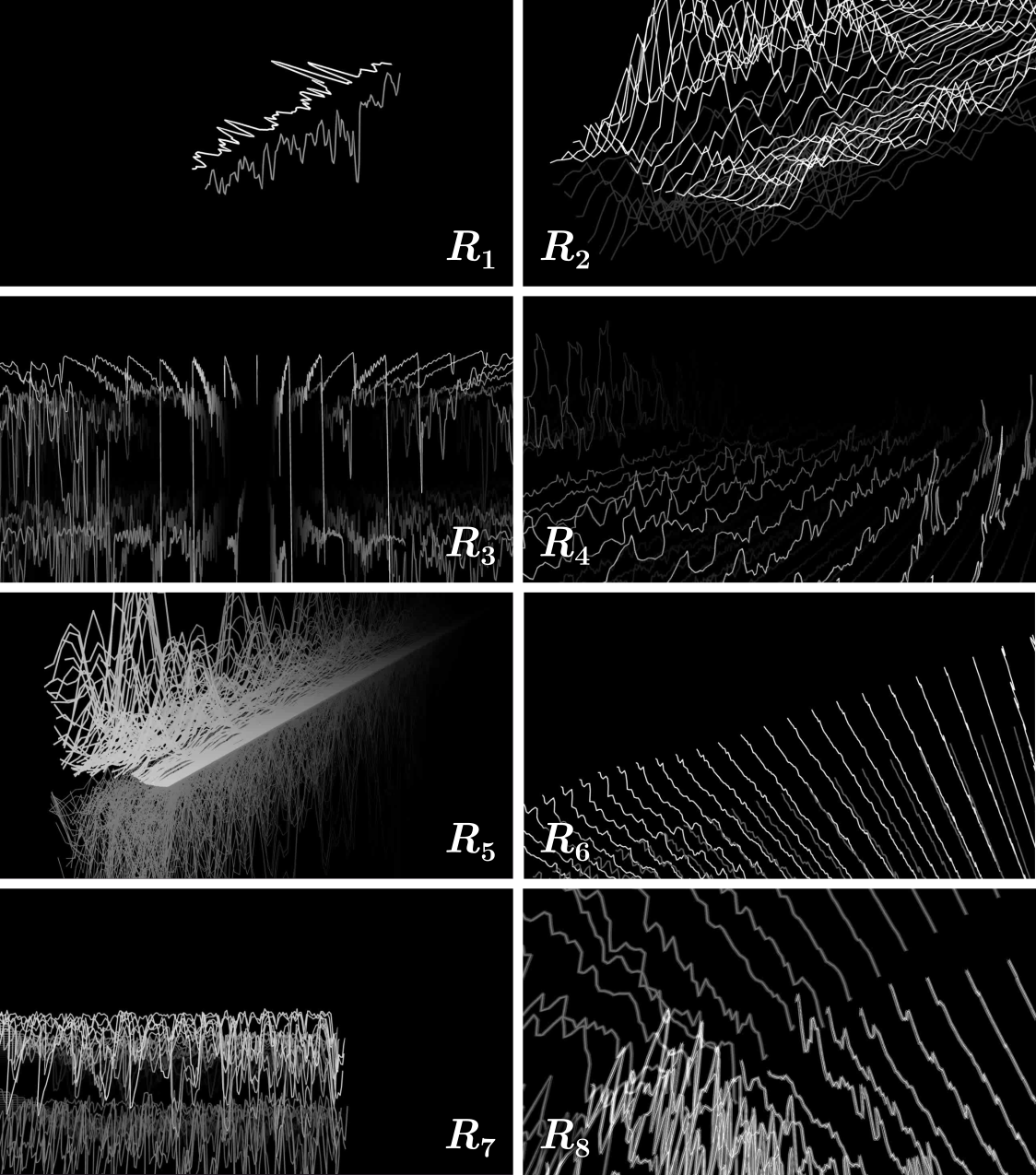}
    \caption{\emph{Visual performance for $8$ Rooms}.  Each panel shows two overlaid audio spectra derived from contact microphones fixed to the motorized rotation stages, while color, line thickness and camera motion are modulated in real time by the corresponding photon-measurement results. Every room has a different accompanying visual performance which are here denoted by $R_1$ to $R_8$.}
    \label{fig:8rooms_visuals}
\end{figure}

Depending on the room being played, the landscape is shown from a different camera angle, and the on-screen positions of Alice and Bob may shift (see Fig.~\ref{fig:8rooms_visuals}). Driven by live measurements, the line behavior is uniquely modified in each of the eight rooms, including changes in thickness, color, spacing, and the application of various visual filters. As a result, each room presents the evolving landscape with its own distinct aesthetic character, mirroring the music’s room-specific sonic identity. Consequently, the visual component of \textit{8 Rooms} not only illustrates the musical form of the composition but also embodies an artistic inquiry into the emergence of patterns within randomness and quantum correlations. Through a minimalist yet expressive visual language rooted in quantum data, it shapes the distinct spatial and expressive atmosphere of each room.

\begin{figure}
    \centering
    \includegraphics[width=1\linewidth]{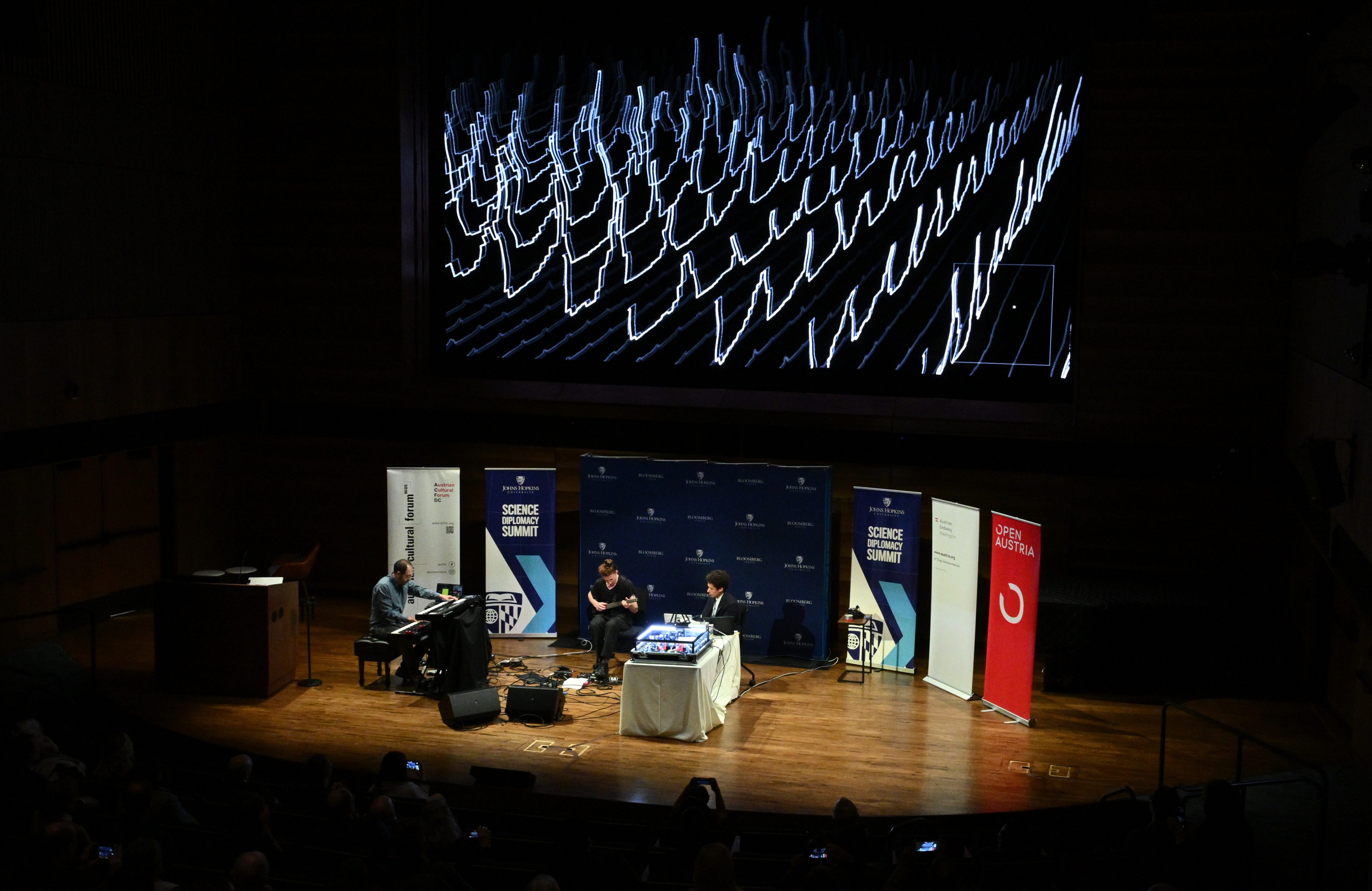}
    \caption{\emph{8 Rooms live performance}. A picture taken during the \textit{$8$ Rooms} performance at the Science Diplomacy Summit 2025 in Washington DC, USA. The Bell setup sits front and center on stage, with the musicians to its side. The visuals for each room are shown on the large screen behind the performers.}
    \label{fig:jhu2025}
\end{figure}

\textit{8 Rooms} premiered on January 25, 2025, at the Vienna Ball of Sciences in Vienna's City Hall \cite{8RoomsLive}. A subsequent performance took place on April 14, 2025, at the Science Diplomacy Summit in Washington D.C., USA (see Fig.~\ref{fig:jhu2025}).

\section{Conclusion}
Quantum physics challenges our understanding of realism and/or locality, opening new pathways for artists and visionaries to explore, interpret and create. The term ``quantum'' in art can be approached in many ways: works may be quantum-inspired, draw on quantum-generated data sets or employ concepts from quantum computing, such as translating an entire octave of musical tones into a single quantum tone~\cite{QuantumMusic}. Or, as in our case, directly use the non-classical correlations from entangled quantum states in a Bell setup. 

\textit{The Sound of Entanglement} represents a synergy between cutting-edge quantum technology and artistic expression. It explores complex scientific principles through sound and visual art, with the entanglement of quantum physics and art serving to familiarize the broader audiences with this emerging field. Our project invites audiences to engage with scientific and artistic experimentation in novel, immersive ways. At the same time, the sonification of experimental data complements the dominance of visual representation and offers us a new means of grasping the vast universe that is quantum physics.

\section{Afterword}

History shows that whenever new technology emerges, art eventually finds a way to explore and embody it. Just as previous advances in electronics and computing transformed music and the visual arts, quantum technologies now inspire new forms of artistic expression. At the same time, artistic vision and creative thinking are essential for the development of science and technology. History also shows that creative approaches encourage exploration beyond conventional frameworks, fostering innovative perspectives and experimentation that often lead to breakthroughs that may not have arisen through traditional analytical methods alone. Consequently, the interplay between art and science enables a more holistic understanding of natural phenomena while inspiring new methodologies and transformative ideas. Today, we are at the forefront of merging a still emerging technology with established art forms to create something new -- something that moves beyond classical frameworks and conventional boundaries.

\section*{Funding Information \& Acknowledgments}
The authors thank the Bruckner Festival 2024, Ars Electronica, and Norbert Trawöger for fruitful discussions, as well as Martina Hochreiter for valuable insights into the history of aleatoric music. The authors also thank the Austrian Cultural Forum Washington DC, especially Verena Daughton and Sarah Bamberger, for creating the opportunity to present this project at the Science Diplomacy Summit 2025. PH thanks the Austrian Science Fund (FWF): Y1121, P36041, P35953, AP and RK also acknowledge financial support from the FWF START Award q-shadows: 713361001. 
This project was supported by the Stadt Wien Kulturabteilung (Arbeitsstipendium Komposition), BMWKMS (Staatsstipendium Komposition), Vienna Center for Quantum Science and Technology (VCQ), FWF SFB BeyondC (10.55776/FG7), Johannes Kepler University Linz, University of Innsbruck, and TU Wien.
\paragraph*{Author contributions:}

 P.H., J.K. and C.W. conceived the concept of The Sound of Entanglement with the help of E.d.D.R., R.K., and M.R.. 
 B.O., M.R. and P.H. designed the experimental Bell setup. B.O. built the setup under the supervision of P.H., with input from all authors, and developed the live communication/transfer of data between the quantum experiment and the musical/visual controls.
 C.W. developed the artistic concept and designed the live performances, with input from all authors.
 E.d.D.R. created the artistic concept of the live visual elements, with input from all authors.
A.P. coordinated the preparation and completion of the manuscript, to which all authors contributed, and produced the documentary.

\onecolumngrid
\section*{References}
\vspace*{-0.8cm}
\twocolumngrid
\bibliography{bibtex}

\clearpage

\section*{Supplementary Material}

\subsection*{Derivation of the CHSH Inequality}

\begin{figure}[t]
\begin{center}
   \begin{tikzpicture}[thick]
    \draw (-1.8,0) node[draw,fill=alicecolor,inner sep=2ex]
    (Alice) {Alice};%
    \draw[black,-*] (Alice.north)+(0,-.2)--+(110:.8);%
    \draw[black,densely dashed,-*] (Alice.north)+(0,-.2)--+(70:.8);%
    \draw[<->,densely dashed]
    (Alice.north)++(0,-.2)+(70:1.1)node[right]{$a_2$}
    arc[radius=1.1,start angle=70,end angle=110] node[left]{$a_1$};%
    \draw[->] (Alice.west)--+(-.5,0.25)node[left]{$A=+1$};
    \draw[->] (Alice.west)--+(-.5,-0.25)node[left]{$A=-1$};%
    \draw (1.8,0) node[draw,fill=bobcolor,inner sep=2ex] (Bob)
    {Bob};%
    \draw[black,-*] (Bob.north)+(0,-.2)--+(110:.8);%
    \draw[black,densely dashed,-*] (Bob.north)+(0,-.2)--+(70:.8);%
    \draw[<->,densely dashed]
    (Bob.north)++(0,-.2)+(70:1.1)node[right]{$b_2$} arc[radius=1.1,start
    angle=70,end angle=110] node[left]{$b_1$};%
    \draw[->] (Bob.east)--+(.5,0.25)node[right]{$B=+1$};
    \draw[->] (Bob.east)--+(.5,-0.25)node[right]{$B=-1$};%
    \draw (0,0) node[draw,circle,fill=sourcecolor,inner sep=1ex]
    (EPR) {Source};%
    \draw[draw=alicecolor,<-] (Alice)--(EPR);
    \draw[draw=bobcolor,->] (EPR)--(Bob);%
  \end{tikzpicture}
\caption{Scheme of the CHSH setup.}
\label{fig:CHSH_scheme}
\end{center}
\end{figure}
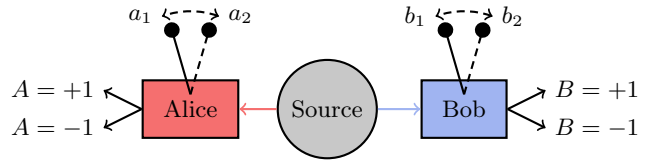

In this section, we try to understand Bell's inequality in a mathematically simplified form, which essentially goes back to John Clauser, Michael Horne, Abner Shimony and Richard Holt (CHSH) \cite{CHSH1969}. To do this, we take a look at Fig.~\ref{fig:CHSH_scheme}. A source emits pairs of photons. In \textit{realism}, every pair is completely determined by (common) hidden variables , which we summarize with the symbol $\lambda$. One of the two photons is sent to a measuring station called Alice, the other to one called Bob. Alice can make one of two possible measurements on her particle, which we denote by $a_1$ and $a_2$. For example, she could control the polarization –- i.e.\ the direction of oscillation of the photon's electric field -- and $a_1$ could be a horizontal (0$^{\circ}$)/vertical (90$^{\circ}$) measurement of polarization, while $a_2$ measures the polarization along a different direction, for example $+45^{\circ}$/$-45^{\circ}$. Similarly, there are two possible measurement settings $b_1$ and $b_2$ for Bob, which do not have to match $a_1$ and $a_2$. We call the result of Alice's measurement $A$ and define that it can only have the values $+1$ and $-1$. For example, if Alice made the horizontal/vertical measurement $a_1$, then $A = +1$ would be the ``horizontal'' result, and $A = -1$ would be the ``vertical'' result. However, if she has taken the $+45^{\circ}$/$-45^{\circ}$ measurement $a_2$, then $A = +1$ would be the result ``$+45^{\circ}$'', and $A = -1$ would be the result ``$-45^{\circ}$''. We define the measurement results $B$ on Bob's side in a similar way.

Since we consider \textit{local} theories with hidden variables, Alice's result $A$ depends, in addition to the hidden variables $\lambda$, only on her setting ($a_1$ or $a_2$) and not on Bob's setting ($b_1$ or $b_2$). Depending on Alice's choice of setting, we name her result $A_1 = A(a_1,\lambda)$ or $A_2 = A(a_2,\lambda)$. According to the example in the previous paragraph, $A_1$ is the result of the horizontal/vertical measurement $a_1$, and $A_2$ is the result of the $+45^{\circ}$/$-45^{\circ}$ measurement $a_2$. Similarly, on Bob's side we name the result $B_1 = B(b_1,\lambda)$ or $B_2 = B(b_2,\lambda)$.

The hidden variables define all four possible measurement results, regardless of whether the measurements are actually carried out or not. In fact, for each pair of photons only either $A_1$ or $A_2$ and only either $B_1$ or $B_2$ can be realized because the respective measurements are mutually exclusive. In the experiment, you have to decide on a polarization direction and cannot measure two different ones at the same time. Since the two measurements $a_1$ and $a_2$ are not compatible (``do not commute''), it is impossible in quantum mechanics to predict both a concrete result for $A_1$ and for $A_2$ for a photon with 100\% probability. The same holds for $B_1$ and $B_2$. However, local realism is precisely the worldview in which even measurements that have not been carried out have definitive values, regardless of what is happening elsewhere at the same time. In theories with local hidden variables, for a single pair of photons all 4 results -- $A_1$, $A_2$, $B_1$, $B_2$ -- have definite values at the same time. Since these can only be $+1$ or $-1$, the following algebraic relation applies:
\begin{align}
    A_1\,(B_1+B_2) + A_2\,(B_1-B_2) = \pm 2. \label{eq:algCHSH}
\end{align}
Here $\pm2$ stands for $+2$ or $-2$. One can check the correctness of this equation by substituting all possibilities for the outcome values $+1$ or $-1$ or convince ourselves as follows: Either $B_1$ and $B_2$ are the same, i.e.\ both are $+1$ or both are $-1$. Then the second bracket is 0 and the first term must be $+2$ or $-2$. Or $B_1$ and $B_2$ are unequal. Then the first bracket is 0 and the result for the second term is $+2$ or $-2$ again.

Now let's imagine that many pairs of particles are created, each with its own hidden variables. In each run, i.e.\ for each individual pair, Alice and Bob randomly choose their settings and thus one of the four possible setting combinations, i.e.\ $(a_1,b_1)$ or $(a_1,b_2)$ or $(a_2,b_1)$ or $(a_2,b_2)$. In around a quarter of the runs they measure $A_1$ and $B_1$, in another quarter $A_1$ and $B_2$ and so on.

In this way, one can experimentally determine all four expected values $E_{11}=\langle A_1B_1\rangle$, $E_{12}=\langle A_1B_2\rangle$, $E_{21}=\langle A_2B_1\rangle$, $E_{22}=\langle A_2B_2\rangle$ of the products of the results -- called correlations. These four numbers are all between $-1$ and $+1$. Since in local realism every single pair must satisfy equality (\ref{eq:algCHSH}), i.e.\ $A_1B_1 + A_1B_2 + A_2B_1 - A_2B_2 = \pm2$, the following inequality applies to the four correlations:
\begin{align}
    S := |E_{11}+E_{12}+E_{21}-E_{22}| \le 2. \label{eq:CHSH}
\end{align}
Simply put: the average of many values that are all $+2$ or $-2$ cannot be greater than +2 and cannot be smaller than $-2$. Inequality (\ref{eq:CHSH}) is a specific version of a Bell inequality, called the CHSH inequality.

All local realist theories must fulfill (\ref{eq:CHSH}). And in contrast to (\ref{eq:algCHSH}), (\ref{eq:CHSH}) can be tested experimentally. Remarkably, in quantum mechanics there are so-called entangled states that predict a violation of Bell's inequality. The polarization singlet state of two photons, for example, produces correlations for which the left side of inequality (\ref{eq:CHSH}) takes the value $S_{\textrm{QM}} = 2\sqrt{2} \approx 2.828$. This quantum state cannot be modeled with local hidden variables. Experiments have confirmed time and time again that the predictions of quantum mechanics are correct and that Bell's inequality is violated in nature, thereby ruling out local hidden variables.

In summary: The worldview in which the properties of physical objects are always well-determined and exist independently of their measurement (realism) and physical influences can only be transmitted at a maximum of the speed of light (locality) contradicts the predictions of quantum physics. This is shown by the experimental violation of Bell's inequality by quantum mechanically entangled states.

\subsection*{2D Random Walk}
A \textit{$2D$ random walk} is a mathematical model which is used to describe random movements on a two dimensional plane, e.g.\ diffusion models in physics. The general idea is that from a starting point on a $2$D lattice, often the point $(x_0,y_0)=(0,0)$, each step leads to a new position $(x_n,y_i)$ with $i=1,2,\ldots$. The general update rule reads:
\begin{eqnarray*}
    x_{n+1} = x_n + s \cos(\theta) \\
    y_{n+1} = y_n + s \sin(\theta)
\end{eqnarray*}
The step size $s$ can be a constant or a (random) function. Each step has its own direction $\theta$, which can be on an orthogonal grid (i.e.\ $\theta \in \{0,\frac{\pi}{2},\pi,\frac{3\pi}{2}\}$) or some other set of angles, or even any direction $\theta \in [0,2\pi)$ in the plane.

In our performance \textit{8 Rooms}, we implemented a special case of such a random walk, where 8 directions are possible, with the angles from the set $\{0, \frac{\pi}{4},\frac{\pi}{2}, \frac{3\pi}{4}, \pi, \frac{5\pi}{4}, \frac{3\pi}{2}, \frac{7\pi}{4}\}$. 
In each measurement run, Alice's and Bob's setting choices $a\in\{a_1,a_2\}$ and $b\in\{b_1,b_2\}$ as well as their outcomes $A\in\{-1,+1\}$ and $B\in\{-1,+1\}$ decide which step in the 2D plane is made. Measurement outcomes performed in $a_1$ or $b_1$ lead to steps along the $x$-axis, and measurements performed in $a_2$ or $b_2$ make steps along $y$-axis. The outcomes decide in which direction along the respective axis the step is made. Concretely, the update rules are:
\begin{eqnarray*}
    x_{n+1}(a,b,A,B) = x_n + A \,\delta_{a,a_1} + B \, \delta_{b,b_1}, \\
    y_{n+1}(a,b,A,B) = y_n + A \, \delta_{a,a_2} + B \, \delta_{b,b_2},
\end{eqnarray*}
with the Kronecker delta function
\begin{equation*}
    \delta_{j,k} = \begin{cases}
    1\, \textrm{if $j$ = $k$}, \\
    0\, \textrm{else}. \\
\end{cases}
\end{equation*}
For example, if settings $a_1,b_2$ are chosen and results $A=+1$, $B=-1$ are obtained, then the random walk update consists of a step along the $+x$ axis and a step down along the $-y$ axis, i.e.\ in total a step along angle $\frac{7\pi}{4}$ is made. If, e.g., Alice and Bob both choose setting 1 and get opposite results, there is no update as the individual contributions cancel.

In total, there are $9$ possible updates for our random walk: the 8 possible directions mentioned above plus the option of ``no step''. In total, there are $16$ setting and outcome combinations, which means some of them yield identical results. For example, the combination $a_1,b_2$ with $A=B=+1$ leads to the same update as the combination $a_2,b_1$ with $A=B=+1$. 

\begin{figure}[t]
    \centering
    \vspace{0.5cm}
\includegraphics[width=0.9\linewidth]{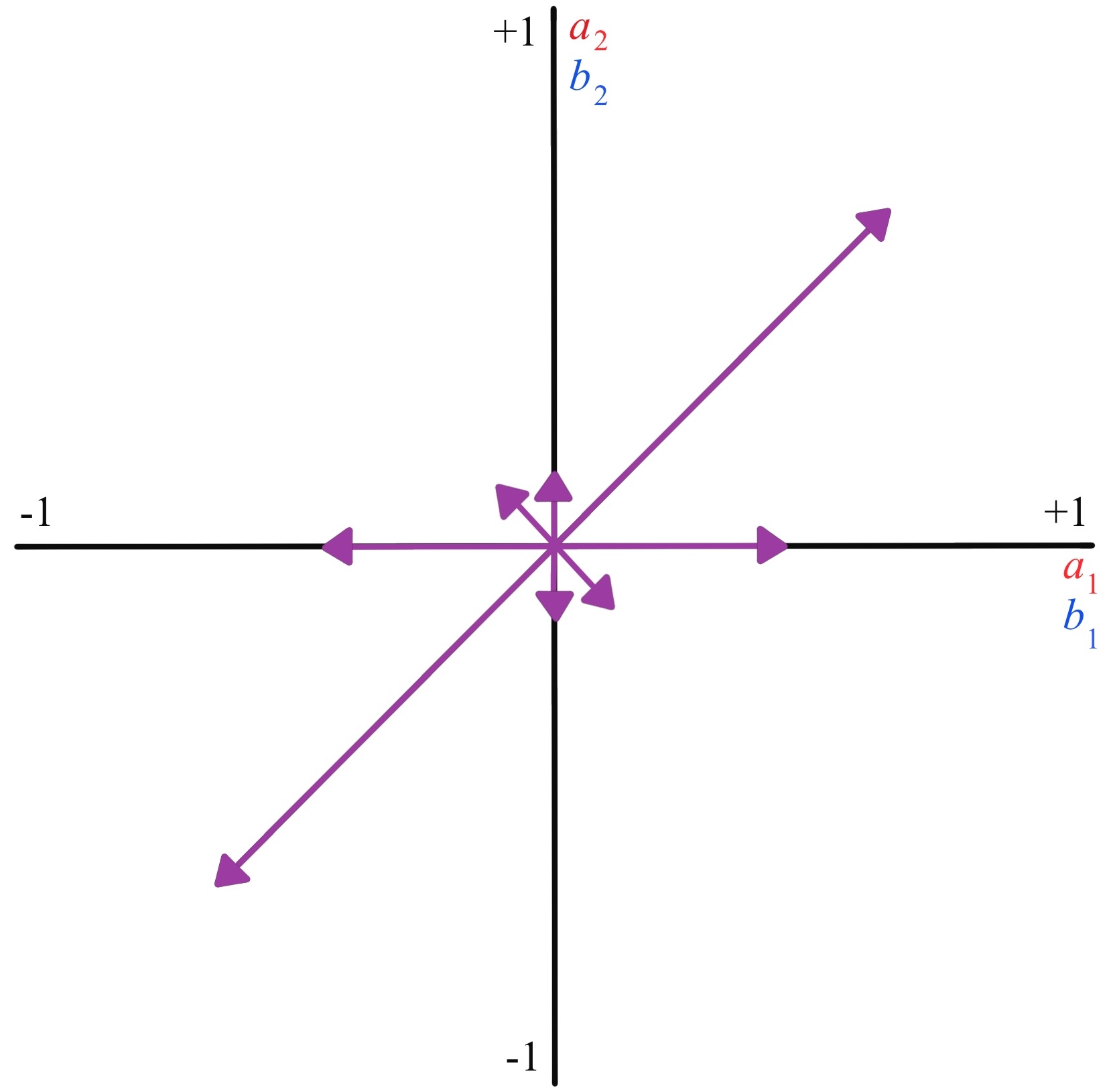}
    \caption{\emph{A skewed quantum random walk with Bell value $2\sqrt{2}$}. This figure shows the projections from $a_1,b_1,a_2,b_2$ onto a two dimensional grid. Measurements yielding $+1$ are translated to arrows going from the center of the grid either right or top, while measurements yielding $-1$ are translated to arrows going from the center of the grid to the left side or down. The length of the arrows are proportional to the amount of times the direction appears as a result of the measurements. When we achieve a Bell value of $2\sqrt{2}$, it is far more likely to go to the upper right hand or lower left hand corner than it is to move along one specific axis. The closer we get to $S=0$ the more the bias vanishes and the probabilities of every direction become equal.}
    \label{fig:qrandomwalk}
\end{figure}

Depending on the correlations of our entangled photons, the random walk can be biased or unbiased. In the case of $S=0$, all directions are equally likely. For a Bell value $0<S\le2$, classical correlations skew the random walk as certain directions are preferred. Quantum correlations stemming from entanglement with $2<S\le2\sqrt{2}$ lead to even larger bias and skewness of the random walk. Fig.~\ref{fig:qrandomwalk} indicates this skewness. 

\end{document}